\documentclass[prl,aps,showpacs,preprint]{revtex4-1}
\usepackage{graphicx}
\usepackage{amsmath}
\usepackage[table]{xcolor}
\usepackage{dcolumn}
\usepackage{multirow}

\begin{document}
\title{Effects of ferroelectric polarization on surface phase diagram: an evolutionary algorithm study of the BaTiO$_{3}$(001) surface}
\author{Pengcheng Chen$^{1,2}$, Yong Xu$^{1,2,3}$, Na Wang$^{1,2}$, Artem R Oganov$^{5,6,7}$, Wenhui Duan$^{1,2,4}$}\email{dwh@phys.tsinghua.edu.cn}
\address{$^1$Department of Physics and State Key Laboratory of Low-Dimensional Quantum Physics, Tsinghua University, Beijing, 100084, People's Republic of China\\
$^2$Collaborative Innovation Center of Quantum Matter, Tsinghua University, Beijing 100084, People's Republic of China\\
$^3$Department of Physics, McCullough Building, Stanford University, Stanford, California 94305-4045, USA\\
$^4$Institute for Advanced Study, Tsinghua University, Beijing 100084, People¡¯s Republic of China\\
$^5$Department of Geosciences, Center for Materials by Design, and Institute for Advanced Computational Science, State University of New York, Stony Brook, NY 11794-2100\\
$^6$Moscow Institute of Physics and Technology, Dolgoprudny city, Moscow Region, 141700, Russian Federation\\
$^7$Northwestern Polytechnical University, Xi'an, 710072, China}
\date{\today}

\begin{abstract}
We have constructed the surface phase diagram of the BaTiO$_{3}$(001) surface by employing an evolutionary algorithm for surface structure prediction, where the ferroelectric polarization is included as a degree of freedom. Among over 1000 candidate structures explored, a surface reconstruction of (2$\times$1)-TiO is discovered to be thermodynamically stable and have the \emph{p2mm} plane group symmetry as observed experimentally. We find that the influence of ferroelectric polarization on the surface free energy can be either negligibly small or sizably large (over 1 eV per ($2 \times 1$) supercell), depending strongly on the surface structure and resulting in a significant distinction of surface phase diagram with varying ferroelectric polarization. It is therefore feasible to control the surface stability by applying an external electric field. Our results may have important implications in understanding the surface reconstruction of ferroelectric materials and tuning surface properties.
\end{abstract} 

\maketitle

The search for stable surface structures is a key subject of surface science and of great importance to fundamental research as well as practical applications, like photovoltaics, catalysis, and sensors\cite{diebold2003surface,young2012first,garrity2013ferroelectric,xu2013space}. Density functional theory (DFT) in combination with \emph{ab initio} thermodynamics is an indispensable tool because of the atomic insight it provides\cite{Oxygen,meyer2004first,eglitis2007ab}. In this approach, the surface free energy is expressed as a function of stoichiometry and atomic chemical potentials so as to consider the varying growth conditions, and the minimization of the surface free energy predicts stable surface phases. Very recently, the approach has been developed for semiconductors to include the electron chemical potential as a new parameter, which can be generally applied to study the effects of bulk dopants on properties of semiconductor surfaces and interfaces\cite{xu2013space}. Further generalization of the approach to other systems would be interesting.

In ferroelectric materials, the ferroelectric polarization couples strongly with the crystal structure, and consequently any change of the ferroelectric polarization will in turn affect the structural stability. Thus, the ferroelectric polarization that can be easily controlled by external electric field is an important degree of freedom for ferroelectric surfaces. However, ferroelectric polarization has hardly been taken into account in the previous calculations of surface phase diagrams of ferroelectric materials. Recent first-principles study indeed showed that surface stability of ferroelectric lithium niobate is different for the positively and negatively polarized surfaces, which is actually driven more by the different surface termination than intrinsic ferroelectric polarization\cite{levchenko2008influence}.

As a prototypical ferroelectric material, barium titanate (BaTiO$_{3}$, BTO) plays a vital role in numerous applications and has been intensively studied theoretically\cite{ramesh2007multiferroics,zheng2004multiferroic,SubSupCond,duan2006predicted,bocher2011atomic,Yujie,Leiye,SM3} As shown in Fig. 1, BTO has a perovskite structure and undergoes a structural transition from high-symmetry cubic phase to low-symmetry tetragonal phase when lowering temperature across $\sim$400 K ~\cite{PhaseDiagram}. The (001) face is a stable cleavage plane and has rich surface reconstructions, including ($1 \times 1$), ($2 \times 1$), $c$($2 \times 2$),($2 \times 2$),($\sqrt{5} \times \sqrt{5}$), ($3 \times 1$), ($3 \times 2$), and ($6 \times 1$) periodicities~\cite{kolpak2008evolution,bando1996structure,JJAP,hudson1993surface,Hagendorf1999106,martirez2012atomic}. Among them, the ($2 \times 1$) reconstruction has recently attracted much attention~\cite{kolpak2008evolution,iles2010systematic,meyerheim2012batio}. Two different surface structure models have been proposed. However, one structure model~\cite{kolpak2008evolution,iles2010systematic} displays the $pm$ plane group symmetry, in contradiction with the $p2mm$ symmetry identified by recent x-ray diffraction experiments~\cite{meyerheim2012batio}; the other one~\cite{meyerheim2012batio} has the correct symmetry but is energetically less stable than the former. The actual atomic structure of the BTO(001)-($2 \times 1$) surface remains elusive.

In this work, we will consider ferroelectric polarization as an extra degree of freedom to calculate surface phase diagram of ferroelectric materials. Specifically, we perform first-principles calculations for the BTO(001) surface, focusing on ($2 \times 1$) as well as ($1 \times 1$) reconstructions to find stable surface configurations and to reveal the effects of ferroelectric polarization. By employing a newly designed evolutionary algorithm\cite{zhu2013evolutionary} for efficiently searching (meta)stable configurations and calculating over 1000 possible structure models, we predict a surface phase diagram containing many new surface structures, including a thermodynamically stable ($2 \times 1$)-TiO phase that has the $p2mm$ plane group symmetry  observed experimentally~\cite{meyerheim2012batio}. More importantly, we find that the influence of ferroelectric polarization on the surface free energy can be either negligibly small or sizably large (over 1 eV per ($2 \times 1$) supercell for BTO(001)), depending strongly on the surface structure. As a result, the surface phase diagram changes significantly with varying ferroelectric polarization. These findings suggest a unique way to control surface structures and properties of ferroelectrics.

\begin{figure}[tbp]
  \centering
  \includegraphics[width=0.6\textwidth]{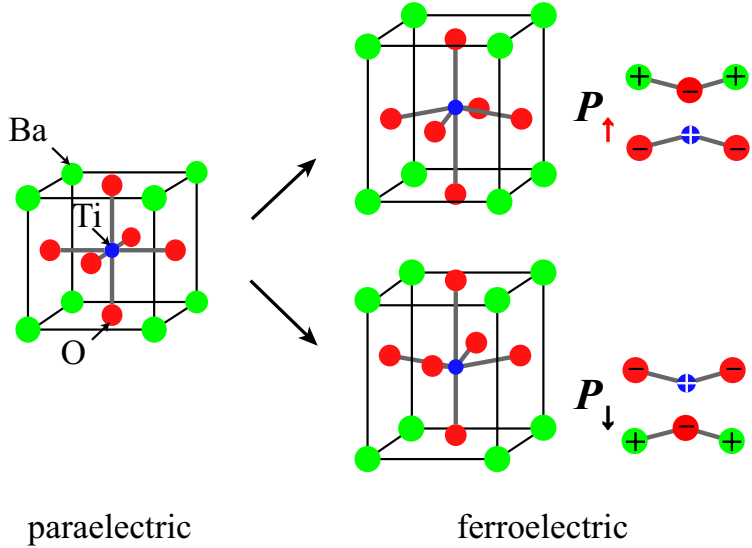}\\
  \caption{(Color online) The atomic structure of bulk BaTiO$_3$ in paraelectric and ferroelectric phase. The polarizations \emph{\textbf{P}}$_{\uparrow}$  and \emph{\textbf{P}}$_{\downarrow}$  are determined by the displacement between O and Ba/Ti along [001]. The green, blue and red balls represent Ba, Ti and O atoms, respectively.}\label{model}
\end{figure}

First-principles calculations were performed with DFT as implemented in the Vienna $\emph{ab initio}$ simulation package\cite{vasp}, using the projector augmented wave method\cite{PAW} and the Perdew-Burke-Ernzerhof (PBE) exchange correlation functional\cite{PBE,NF}. The method for predicting surface reconstructions was based on an evolutionary algorithm as implemented in the USPEX package\cite{oganov2006crystal}, allowing variable-composition structure searches, where the numbers of atoms in the surface region are varied to yield the global minimum of the surface free energy (see Refs. \cite{zhu2013evolutionary,ZhouXF,WangQG} for more details). To demonstrate the effect of ferroelectric polarization on surface reconstructions, we focused on discussing the TiO$_{2}$-terminated surfaces, which have been extensively observed experimentally and studied theoretically\cite{kolpak2008evolution,martirez2012atomic,iles2010systematic,martirez2012atomic}, and considered only an ideal bulk-terminated phase for the BaO-terminated surface.  The BTO(001) surfaces were modeled by periodic slabs composed of four -TiO$_{2}$-BaO- bilayers plus a TiO$_{2}$ termination together with a 15 \AA \ thick vacuum layer\cite{NS}. Different surface stoichiometries were considered by adding a layer of Ti$_{x}$O$_{y}$ ($x$=0, 1, 2, $y$=0, 1, 2, 3, 4) in a ($2 \times 1$) surface supercell on the otherwise ideal TiO$_{2}$-terminated surface. The bottom three bilayers were fixed at their bulk configuration, and the other layers were relaxed using the conjugate gradient algorithm till residual forces were smaller than 0.01 eV/\AA$^{2}$. A Monkhorst-Pack $k$-point mesh with reciprocal-space resolution of 2 $\pi$ $\times$ 0.03 \AA$^{-1}$ and a 400 eV planewave cutoff energy were used. Dipole correction was employed in slab calculations for removing artificial interactions between the slab and its periodic images.

Thermodynamical stability of a surface structure is determined by the surface free energy, $\gamma  = {G_{\rm slab}} - {G_{\rm ref}} - \Delta {n_{\rm Ba}}{\mu _{\rm Ba}} - \Delta {n_{\rm Ti}}{\mu _{\rm Ti}} - \Delta {n_{\rm O}}{\mu_{\rm O}}$, where $G_{\rm slab}$ and $G_{\rm ref}$ are the Gibbs free energies of the slab and the reference system that was selected as the ideal TiO$_{2}$-terminated surface. $\Delta {n_{\rm Ba}}$, $\Delta {n_{\rm Ti}}$ and $\Delta {n_{\rm O}}$ denote the changes in the number of atoms with respect to the reference system. All these quantities correspond to a ($2 \times 1$) surface supercell if not specified otherwise. $\mu_{\rm Ba}$, $\mu_{\rm Ti}$  and $\mu_{\rm O}$ are the atomic chemical potentials. The accessible boundary of chemical potentials is defined by thermal equilibria between bulk BaTiO$_{3}$ and other phases, including bulk Ba, bulk Ti, bulk BaO and bulk TiO$_{2}$. Herein we approximated the Gibbs free energy by the DFT total energy, excluding the vibrational contribution. The approximation has been found to be satisfactory for our study: the phase diagram qualitatively remains unchanged when the temperature effect was considered (see the Supplemental Material~\cite{SM}).

At first we exclude the contribution of ferroelectric polarization by fixing the lower three -TiO$_{2}$-BaO- bilayers at the cubic bulk structure (i.e., paraelectric phase), as typically done in previous studies~\cite{iles2010systematic,kolpak2008evolution,meyerheim2012batio}. In contrast to previous studies, we computed more surface configurations (over 1000) using an advanced evolutionary algorithm\cite{zhu2013evolutionary}, and obtained many new stable surface structures. Figure 2(a) shows the computed surface phase diagram of BTO(001) for ($1 \times 1$) and ($2 \times 1$) reconstructions, and Fig. 2(b) shows the atomic configurations of the stable phases. It can be seen from Fig. 2(a) that the ($1 \times 1$) ideal BaO-terminated surface is stable at O-rich and Ba-rich conditions. With decreasing $\mu_{\rm O}$ and $\mu_{\rm Ba}$, other phases become increasingly more stable. These stable phases include a double-layer TiO$_2$-termination model ($2 \times 1$)-Ti$_{2}$O$_{4}$~\cite{kolpak2008evolution,iles2010systematic}, a TiO adunit model ($2 \times 1$)-TiO formed by adding a TiO unit vertically at the hollow site, and two Ti adatom models, ($2 \times 1$)-Ti and ($1 \times 1$)-Ti, formed by adding a Ti atom at the hollow site in the surface supercell. Note that another double-layer model ($1 \times 1$)-TiO$_{2}$ in Fig. 2(b) (not shown in the phase diagram) has surface free energy very close (within $\sim$ 2 meV/\AA$^2$) to that of the ($2 \times 1$)-Ti$_{2}$O$_{4}$ phase. Similar geometrical features of Ti$=$O titanyl are found in these two thermodynamically degenerate phases.

\begin{figure}[tbp]
  \centering
  \includegraphics[width=0.8\textwidth]{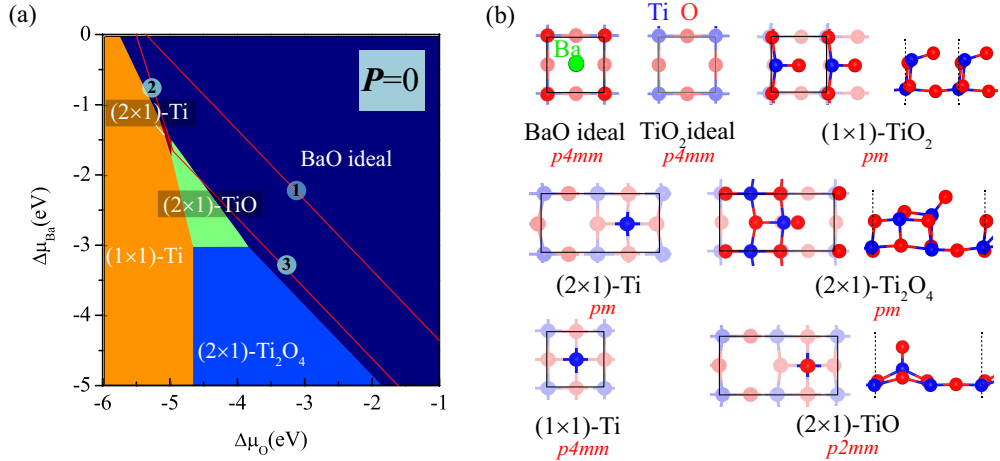}\\
  \caption{(Color online) (a) Surface phase diagram of paraelectric BTO(001) for ($1 \times 1$) and ($2 \times 1$) reconstructions without the effect of ferroelectric polarization. The red solid lines bound the chemical stability ranges of BTO. The precipitation lines of bulk BaO, Ti and TiO$_{2}$ are labeled as 1, 2 and 3, respectively, which bound the accessible chemical potential range defined by thermal equilibria. (b) Schematic atomic structures of stable surface phases. The blue and red balls represent Ti and O atoms, respectively. }\label{phasediagram}
\end{figure}
%as a function of the relative chemical potentials of Ba ($\Delta \mu_{\texttt{Ba}}$) and O ($\Delta \mu_{\texttt{Ba}}$)

In experiment, the BTO(001) ($2 \times 1$) surface reconstruction is obtained by Ar$^+$ ion sputtering and subsequently annealing native ($1 \times 1$) surface~\cite{meyerheim2012batio}. Our surface phase diagram suggests three ($2 \times 1$) phases, including the known double-layer model ($2 \times 1$)-Ti$_{2}$O$_{4}$~\cite{kolpak2008evolution,iles2010systematic}. Another double-layer ($2 \times 1$) model proposed by previous work~\cite{meyerheim2012batio} does not appear in the phase diagram for the reason that it is thermodynamically less stable and its surface free energy is far higher ($\sim$ 1.4 eV) than that of the double-layer model ($2 \times 1$)-Ti$_{2}$O$_{4}$ shown in the phase diagram. Careful analysis of X-ray diffraction data indicates that the ($2 \times 1$) surface has the $p2mm$ plane group symmetry~\cite{meyerheim2012batio}. Based on this information, the ($2 \times 1$)-Ti$_{2}$O$_{4}$ and ($2 \times 1$)-Ti models that have the $pm$ symmetry are excluded. The only remaining ($2 \times 1$) model with the $p2mm$ symmetry is ($2 \times 1$)-TiO, which is a thermodynamically stable phase located within the chemical stability ranges of BTO [see Fig. 2(a)]. Interestingly, among over 1000 structures suggested by the evolutionary algorithm, the ($2 \times 1$)-TiO model is the only one that satisfies both conditions of energy and symmetry. We thus attribute the formation of ($2 \times 1$) reconstruction to the addition of TiO units, which is consistent with recent experimental observation of TiO adunits on the $c$(2$\times$2) BTO(001) surface~\cite{Kinetics}.

We then consider different ferroelectric polarizations (\emph{\textbf{P}}$_{\downarrow}$ and \emph{\textbf{P}}$_{\uparrow}$) by fixing the lower three -TiO$_{2}$-BaO- bilayers  at the tetragonal bulk structures (ferroelectric phase) as shown in the right panel of Fig. 1. Figure 3 shows the calculated surface phase diagram for two opposite ferroelectric polarization orientations. The interesting (2$\times$1)-TiO structure remains as a stable phase in the surface phase diagram for both types of polarizations. However, distinct variations of phase diagrams in different polarization conditions can be found, e.g., (2$\times$1)-TiO is stable in the \emph{\textbf{P}}$_{\downarrow}$ condition over a remarkably extended chemical potential range, while under the \emph{\textbf{P}}$_{\uparrow}$ condition (2$\times$1)-TiO becomes unstable unless the $\mu$$_{O}$ is fairly low. The distinction between the phase diagrams under different polarizations indicates that at a certain experimental circumstance (e.g., the oxygen chemical potential is within $-3$ eV $\sim$ $-5$ eV), the thermodynamically stable phases are different (i.e., (2$\times$1)-TiO and (1$\times$1) ideal BaO-terminated surfaces). Thus the external electric field induced ferroelectric switching of the substrate might lead to appearance of different surface structures if kinetic factors were not considered.

\begin{figure}[tbp]
  \centering
  \includegraphics[width=0.8\textwidth]{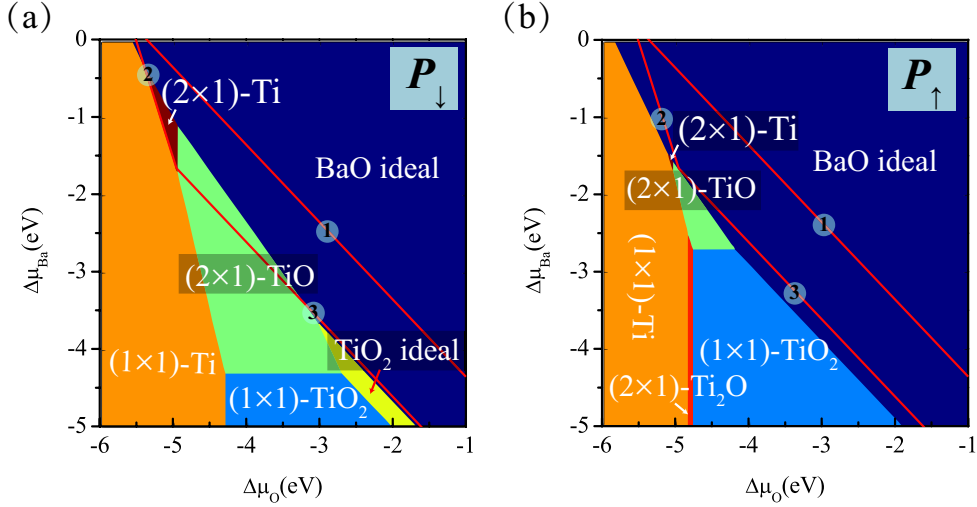}\\
  \caption{(Color online) Same as Fig. 2(a) except that the effect of ferroelectric polarization is included. Different ferroelectric polarizations, (a) \emph{\textbf{P}}$_{\downarrow}$ and (b) \emph{\textbf{P}}$_{\uparrow}$ as illustrated in Fig. 1(b), result in significantly different surface phase diagrams. }\label{figure3}
\end{figure}

To understand this phenomenon, we quantitatively analyze the influence of ferroelectric polarization on the phase diagram by calculating the relative surface Gibbs free energies,
$\Delta$$\gamma$($\emph{\textbf{P}}_{\downarrow/\uparrow}$)=$\gamma$($\emph{\textbf{P}}_{\downarrow/\uparrow}$)$-$$\gamma$({\emph{\textbf{P}}$_{=0}$}), for various surface structures under two opposite polarization conditions (the case without ferroelectric polarization is taken as the reference). As shown in Table 1, the obtained nonstoichiometric reconstruction phases can be divided into two types: the ones with a TiO$_{2}$ overlayer and the other ones with a Ti$_{x}$O$_{y}$ ($y$$<$$2x$) adunit on the primary TiO$_{2}$ termination. The surfaces of the ``adunit''-type show a considerable energy difference of $\Delta\gamma(\emph{\textbf{P}}_{\downarrow})- \Delta\gamma(\emph{\textbf{P}}_{\uparrow})$  (~$\sim$ 1.0 eV/(2$\times$1)cell), indicating significant influence of the ferroelectric polarization on the surface stability. In contrast, the corresponding influence is much smaller for the ``overlayer''-type surfaces. Our results of structural relaxation explicitly show that the detailed surface atomic structures of stable phases hardly change with the effect of ferroelectric polarization. Thus, the remarkable difference between $\Delta$$\gamma$($\emph{\textbf{P}}_{\downarrow}$) and $\Delta$$\gamma$($\emph{\textbf{P}}_{\uparrow}$)  does not come from structural relaxation but is mainly caused by electrostatic interactions.

Generally, termination of the spontaneous polarization of ferroelectric materials always gives rise to discontinuity of polarization at surfaces, leading to surface polarization charges (and surface metallicity) whose signs depend on the direction of polarization (see more details in Supplementary Material\cite{SM}). As shown in Figs. 4(a) and 4(b), these surface polarization charges generate a depolarization field (or an internal electric field), whose direction is opposite to that of the polarization. With a constant non-zero depolarization field, the electrostatic energy of ferroelectric surfaces would diverge with increasing thickness. Such an electrostatic instability, however, can be eliminated by compensating the depolarization field through various mechanisms, like introducing an external electric voltage\cite{SM2,SM4}, surface adsorption or surface reconstruction. For the intrinsic mechanism of surface reconstruction, the compensation of the depolarization field depends significantly on the direction of polarization, leading to distinct surface free energies $\Delta$$\gamma$($\emph{\textbf{P}}_{\downarrow}$) and $\Delta$$\gamma$($\emph{\textbf{P}}_{\uparrow}$).

Different surface reconstructions result in different surface electrostatic potentials and thus correspond to varying surface dipoles. For a specified surface reconstruction, if the depolarization field is compensated by the surface dipole [see Figs. 4(c) and 4(d)], the electrostatic energy would get lowered by this compensation and the whole system thus gets stabilized. In contrast, reversing the depolarization field would yield higher electrostatic energy and larger surface free energy. This could well explain the polarization dependent behaviors, as demonstrated below. While the magnitude of surface dipole is not easy to quantify, a qualitative analysis is possible for the present system, considering that the normal oxidation states of Ti and O are $+4$ and $-2$, respectively.

\begin{table}
\caption{\label{Tab 1} Relative surface free energies [$\Delta \gamma $] and Bader charges for the surface adunits on the TiO$_{2}$ termination for the non-stoichiometric phases in different polarization conditions. The case without ferroelectric polarization is taken as the reference of $\Delta \gamma $.}
\begin{ruledtabular}
\begin{tabular}{lcccc}
\multirow{2}{*}{Phase} &\multicolumn{2}{c}{$\Delta$$\gamma$(eV)}&\multicolumn{2}{c}{Charge(e)}\\\cline{2-3}\cline{4-5}
& $\textbf{P}_{\downarrow}$ & $\textbf{P}_{\uparrow}$ & $\textbf{P}_{\downarrow}$ & $\textbf{P}_{\uparrow}$ \\ \hline
(2$\times$1)-Ti$_{2}$O$_{4}$  & 0.47   &     0.02    &0.03  & 0.02   \\
(1$\times$1)-TiO$_{2}$        & 0.38   &     $-0.09$ &0.06  &$-0.02$ \\
(2$\times$1)-TiO              &$-0.82$ &	 0.32    &1.18  & 1.15   \\
(2$\times$1)-Ti               &$-0.87$ &	 0.37    &1.46  & 1.45   \\
(1$\times$1)-Ti               &$-1.00$ &	 0.48    &2.58  & 2.59   \\
(2$\times$1)-Ti$_{2}$O        &$-0.77$ &	 0.17    &1.50  & 1.53   \\
\end{tabular}
\end{ruledtabular}
\end{table}

For the so-called ``adunit''-type surfaces, the adunit that has Ti/O ratio larger than 1/2 is chemically unsaturated. When the adunit binds with the substrate, electron transfer from the adunit to the substrate occurs, resulting in a positively charged adunit as shown in Fig. 4(c). For the \emph{\textbf{P}}$_{\downarrow}$ condition, the charge transfer decreases the surface polarization charge and the charge-transfer induced dipole compensates the depolarization field, resulting in a negative $\Delta$$\gamma$($\emph{\textbf{P}}_{\downarrow}$). We denote such a charge-transfer induced compensation as the ionic surface compensation mechanism, as used in previous work~\cite{IonicSurfaceCompensation}. The effect gets inverted for the \emph{\textbf{P}}$_{\uparrow}$ condition, leading to a positive $\Delta$$\gamma$($\emph{\textbf{P}}_{\uparrow}$). All these features are consistent with the calculation data (see Table 1). To analyze the results in more details, we present the calculated Bader charges in Table 1. Due to the lower coordination number of surface atoms, the calculated charge of the adunits, e.g., TiO, which have formed Ti=O double-bonded titanyl groups, is about +1.18e. A large surface free energy difference of ``adunit''-type phases [$>$ 1.0 eV/(2$\times$1)cell] is induced through the ionic surface compensation mechanism described above. The detailed spatial distribution of charge transfer is given in the Supplementary Material\cite{SM}.

\begin{figure}[tbp]
  \centering
  \includegraphics[width=0.8\textwidth]{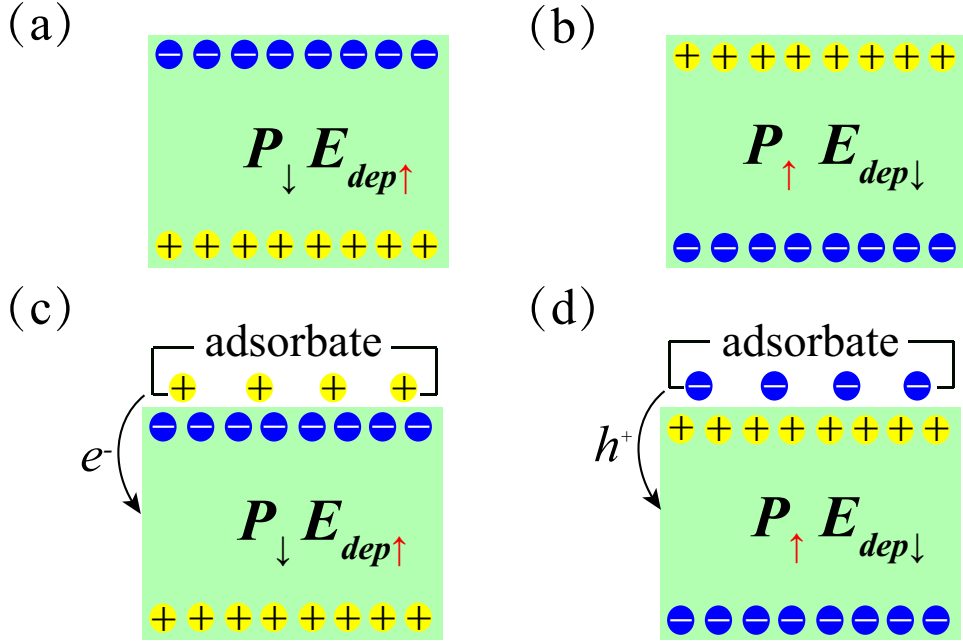}\\
  \caption{(Color online) (a)/(b) The discontinuity of ferroelectric polarization (\emph{\textbf{P}}$_{\downarrow}$/\emph{\textbf{P}}$_{\uparrow}$) at the surface induces surface polarization charges, resulting in a depolarization field opposite to the direction of polarization. (c)/(d) Schematic of ionic surface compensation mechanism. Charge transfer between the adsorbate and substrate induces a surface dipole, which compensates the depolarization field and thus lowers the electrostatic energy of the system, leading to lower surface free energy.}\label{figure4}
\end{figure}

For the so-called ``overlayer''-type phases in which the overlayer itself is chemically saturated, there exists tiny charge transfer between the overlayer and the substrate (see Table 1), suggesting that the effect of ionic surface charge compensation is negligible. The atomic rumpling of TiO$_{2}$ overlayer, with the O atoms at the surface all above the Ti atoms, contributes a downward surface dipole \emph{\textbf{P}}$_{s}(\downarrow)$. Such a kind of structural rumpling has also been predicted for the bare surfaces of perovskites which can lead to a relatively low catalytic activity of the surface\cite{fechner2008effect}. The anti-parallel/parallel configuration of the \emph{\textbf{P}}$_{s}$ to the \emph{\textbf{P}}$_{\uparrow}$ and \emph{\textbf{P}}$_{\downarrow}$ of the substrate leads to the stabilization/destabilization scenarios, respectively. Thus, the polarization to the surface is suppressed for the \emph{\textbf{P}}$_{\uparrow}$ condition but enhanced for \emph{\textbf{P}}$_{\downarrow}$ condition, resulting in the higher $\Delta$$\gamma$. The resulting changes of electrostatic potential alignment and the surface electronic structure by different types of reconstruction are given in the Supplementary Material\cite{SM}. The obtained (2$\times$1)-TiO surface shows \emph{n} type metallicity in both polarization conditions.

It should be noticed that the contributions of charge transfer and structural rumpling are strongly entangled and cannot be clearly distinguished by direct calculations, especially in the ``adunit''-type phases like (2$\times$1)-TiO. Nevertheless, compared to the relatively smaller [$\Delta$$\gamma$($\emph{\textbf{P}}_{\downarrow}) - \Delta$$\gamma$($\emph{\textbf{P}}_{\uparrow}$)] of (1$\times$1)-TiO$_{2}$ case ($\sim$ 0.4 eV) which is basically contributed by the structural rumpling, the considerably larger [ $\Delta$$\gamma$($\emph{\textbf{P}}_{\downarrow}) - \Delta$$\gamma$($\emph{\textbf{P}}_{\uparrow}$)] of (2$\times$1)-TiO case ($\sim$ $-1.1$ eV) can be mainly attributed to charge transfer (i.e., the ionic surface compensation mechanism).

The above results as well as related physical mechanism clearly indicate that the ferroelectric polarization plays a significant role in the surface stability of ferroelectric materials. This reveals a new degrees of freedom to affect the growth of the surface: in addition to tuning the growth condition, e.g., substrate temperature and partial pressure of source, it is convenient and feasible to control the surface stability by applying an external electric field.

In summary, we have constructed the surface phase diagram of (2$\times$1) and (1$\times$1) BTO(001) reconstructions by employing a surface structure prediction method based on evolutionary algorithm and exploring over 1000 candidate structures. We predict a surface phase diagram containing many new surface structures, including a thermodynamically stable (2$\times$1)-TiO phase that has the \emph{p2mm} plane group symmetry as observed experimentally. Critically, the ferroelectric polarization has been included as a new parameter of surface structure prediction. We find the surface phase diagram changes significantly with varying ferroelectric polarization due to the ionic surface compensation mechanism. The distinguishing feature of ferroelectrics is the polarization switching upon applying external electric field, thus the control over surface stability is feasible by applying electric field. The underlying physical mechanism is expected to be quite general. Our results may help in tuning surface structures and properties of ferroelectric materials.

We thank Zhirong Liu and Xiangfeng Zhou for valuable discussions. We acknowledge the support of the Ministry of Science and Technology of China (Grant Nos. 2011CB921901 and 2011CB606405), and the National Natural Science Foundation of China. A.R.O. thanks the National Science Foundation (EAR-1114313, DMR-1231586), DARPA (Grants No. W31P4Q1210008 and No. W31P4Q1310005), the Government (No. 14.A12.31.0003) and the Ministry of Education and Science of Russian Federation (Project No. 8512) for financial support, and Foreign Talents Introduction and Academic Exchange Program (No. B08040).

\providecommand{\noopsort}[1]{}\providecommand{\singleletter}[1]{#1}%


\begin{thebibliography}{41}
\bibitem [{NS()}]{NS}%
  \BibitemOpen
  \href@noop {} {}\bibinfo {note} {The selection of the bottom surface
  termination (i.e., TiO$_{2}$- or BaO- terminated bottom surface) has been
  found not to affect the results. We have also performed test calculations on
  the slab thickness effect by adding extra unit cell to the slab, which yield
  very close relative surface free energies and the same surface
  stability.}\BibitemShut {Stop}%
\end{thebibliography}
\end{document}